\newcounter{myctr}
\def\myitem{\refstepcounter{myctr}\bibfont\noindent\ifnum\themyctr>9\else\phantom{0}\fi\hangindent17pt\t2hemyctr.\enskip}

\documentclass{ws-ijqi}

\newcommand{\mathpi}{\pi}
\newcommand{\mathd}{\mathrm{d}}
\newcommand{\mathe}{\mathrm{e}}

\newcommand{\tmop}[1]{\ensuremath{{#1}}}
\newcommand{\tmtextbf}[1]{{\bfseries{#1}}}
\newcommand{\tmtextit}[1]{{\itshape{#1}}}

\usepackage{amsmath}
\usepackage{amssymb}
\usepackage{amscd}
\usepackage{bbm}

\begin{document}

\markboth{B. Vacchini}
{General structure of quantum collisional models}

\catchline{}{}{}{}{}

\title{General structure of quantum collisional models}

\author{Bassano Vacchini}

\address{Dipartimento di Fisica, Universit{\`a} degli Studi di
Milano, Via Celoria 16, I-20133 Milan, Italy\\
INFN, Sezione di Milano, Via Celoria 16, I-20133
Milan, Italy\\
bassano.vacchini@mi.infn.it}

\maketitle

\begin{history}
\received{\today}
\end{history}

\begin{abstract}
We point to the connection between a recently introduced class of
non-Markovian master equations and the general structure of quantum
collisional models. The basic construction relies on three basic ingredients:
a collection of time dependent completely positive maps, a completely
positive trace preserving transformation and a waiting time distribution
characterizing a renewal process. The relationship between this construction
and a Lindblad dynamics is clarified by expressing the solution of a Lindblad
master equation in terms of demixtures over different stochastic trajectories
for the statistical operator weighted by suitable probabilities on the
trajectory space.
\end{abstract}


\section{Introduction}

The study of open quantum systems has a long history
{\cite{Davies1976,Breuer2007}}, and an important and difficult topic from the
very beginning was the treatment of non-Markovian dynamical evolutions. Such
dynamics should describe memory effects, which typically appear in the
presence of strong system environment interaction, at low temperature, or if
the environment has a complex spectral density. Recently new important
results have been obtained in the very definition of a non-Markovian dynamics
{\cite{Breuer2009b,Rivas2010a,Wolf2008a}}, see e.g. {\cite{Breuer2012a}} for a
first summary of recent results.

A well know class of open system dynamics, which is considered to be Markovian
whatever definition of non-Markovianity one adopts, even the more strict and
close to the classical notion as discussed in
{\cite{Lindblad1979a,Dumcke1983a}}, is given by the generator of a completely
positive quantum dynamical semigroup as obtained by Gorini, Kossakowski,
Sudarshan and Lindblad {\cite{Gorini1976a,Lindblad1976a}}. This result,
besides providing a quite general class of well defined time evolutions for a
system interacting with some environment, has many important special features
{\cite{Breuer2007,Rivas2012}}. Among others, its operator structure can be
naturally related to elementary physical interactions, and the solution of the
master equation can be related to a measurement interpretation, arising as an
average over suitable stochastic realizations depending on the measurement
outcomes. A major effort in the description of open quantum systems is the
quest for a generalization of this class of Markovian time evolutions,
possibly keeping some of its nice features. The minimal requirement is the
preservation of trace and positivity of the statistical operator, which is
granted by complete positivity of the time evolution. However other
interesting features in looking for extensions are given by a connection
between the operator structure of the equation and physically relevant or
addressable quantities, as well as a possible link to a realization of the
overall dynamics in terms of simpler and experimentally more manageable
evolutions. This is also one of the motivations behind the so called
collisional models
{\cite{Ziman2005a,Rybar2012a,Giovannetti2012a,Ciccarello2013a}}, which
describe the dynamics as the effect of repeated interactions with an
environment, considered as isolated collisions.

\section{Piecewise dynamics from Lindblad equation}

In this paper we want to point out in which sense a recent result about a
general class of non-Markovian time evolutions {\cite{Vacchini2013a}} can be
seen as a generalization of the Lindblad master equation. That is to
say in which sense it can be seen as a structure of master
equation, whose solutions are warranted to provide a completely
positive time evolution map, and which includes a semigroup evolution
as a special case.
We will see that it
should more properly be seen as the general structure of a collisional model
describable in terms of a closed evolution equation for the statistical
operator of the system only. To this aim we consider a particular expression
of the solution of the Lindblad dynamics, which opens the way for a general
characterization of certain collisional models. Given an equation of the form
\begin{eqnarray}
  \frac{\mathd \rho}{\tmop{dt}} & = & \mathcal{L} \rho,  \label{eq:lindblad}
\end{eqnarray}
where $\rho$ denotes the statistical operator describing the open system and
$\mathcal{L}$ is a linear superoperator, it has been shown that the ensuing
dynamics is given by a semigroup of completely positive time evolutions if and
only if the superoperator is of the form {\cite{Gorini1976a,Lindblad1976a}}
\begin{eqnarray}
  \mathcal{L} \rho & = & \mathcal{L}_R \rho + \mathcal{J} \rho, \nonumber
\end{eqnarray}
where
\begin{eqnarray}
  \mathcal{L}_R \rho & = & R \rho + \rho R^{\dag} \nonumber\\
  R & = & - \frac{i}{\hbar} H - \frac{1}{2} \sum_k L_k^{\dag} L_k, \nonumber
\end{eqnarray}
with $H$ a self-adjoint operator, and $\mathcal{J}$ denotes the completely
positive superoperator
\begin{eqnarray}
  \mathcal{J} \rho & = & \sum_k L_k \rho L_k^{\dag} .  \label{eq:l}
\end{eqnarray}
Beside the completely positive map $\mathcal{J}$ it is natural to introduce
the semigroup of completely positive trace decreasing superoperators
$\mathcal{R} \left( t \right)$
\begin{eqnarray}
  \mathcal{R} \left( t \right) \rho & = & \mathe^{tR} \rho \mathe^{tR^{\dag}}
  .  \label{eq:r}
\end{eqnarray}
Together with the latter it is convenient for our purposes to introduce the
following superoperators, which send positive operators to statistical
operators
\begin{eqnarray}
  \mathcal{\tilde{J}} \sigma & = & \frac{\mathcal{J} \sigma}{\tmop{Tr} \left(
  \mathcal{J \sigma} \right)}  \label{eq:jtilde}
\end{eqnarray}
and
\begin{eqnarray}
  \tilde{\mathcal{R}} \left( t \right) \sigma & = & \frac{\mathcal{R} \left( t
  \right) \sigma}{\tmop{Tr} \left( \mathcal{R} \left( t \right) \sigma
  \right)} .  \label{eq:rtilde}
\end{eqnarray}
Note that the latter transformations, denoted by a tilda, are not linear, but
rather homogeneous of order zero according to the relation
\begin{eqnarray}
  \mathcal{\widetilde{\mathcal{A}}} \left( \mu \sigma \right) & = &
  \mathcal{\tilde{A}} \sigma \hspace{2em} \forall \mu \in \mathbbm{C} 
  \label{eq:homo}
\end{eqnarray}
which holds for both $\mathcal{\tilde{J}}$ and $\tilde{\mathcal{R}} \left( t
\right)$, so that in particular if $\sigma$ is itself a statistical operator
they can be seen as repreparations of the state according to the action of the
map $\mathcal{J}$ or $\mathcal{R} \left( t \right)$ respectively. In terms of
these operators the solution of Eq.~(\ref{eq:lindblad}) can be expressed as
follows
\begin{eqnarray}
  \rho \left( t \right) & = & \mathpi^0_t \left( t \right) \tilde{\mathcal{R}}
  \left( t \right) \rho \left( 0 \right) + \sum_{n = 1}^{\infty} \int^t_0
  \mathd t_n \ldots \int^{t_2}_0 \mathd t_1 \mathpi^n_t \left( t_1, \ldots,
  t_n \right)  \label{eq:physprob}\\
  &  & \hspace{2em} \times \mathcal{\tilde{\mathcal{R}}} \left( t - t_n
  \right) \mathcal{\tilde{J}} \ldots \mathcal{\widetilde{J}} \mathcal{R}
  \left( t_2 - t_1 \right) \widetilde{\mathcal{J}} 
  \mathcal{\widetilde{\mathcal{R}}} \left( t_1 \right) \rho \left( 0 \right),
  \nonumber
\end{eqnarray}
where besides the operators $\mathcal{\tilde{J}}$ and $\widetilde{\mathcal{R}}
\left( t \right)$ defined in Eq.~(\ref{eq:jtilde}) and Eq.~(\ref{eq:rtilde})
respectively, we have defined the quantity
\begin{eqnarray}
  \mathpi^0_t \left( t \right) & = & \tmop{Tr} \left( \mathcal{R} \left( t
  \right) \rho \left( 0 \right) \right)  \label{eq:po}
\end{eqnarray}
which can be interpreted as the probability of no jumps described by the
superoperator $\mathcal{J}$ up to time $t$, while
\begin{eqnarray}
  \mathpi^n_t \left( t_1, \ldots, t_n \right) & = & \tmop{Tr} \left(
  \mathcal{R} \left( t - t_n \right) \mathcal{J} \ldots \mathcal{J}
  \mathcal{R} \left( t_2 - t_1 \right) \mathcal{J} \mathcal{R} \left( t_1
  \right) \rho \left( 0 \right) \right),  \label{eq:pn}
\end{eqnarray}
can be taken as exclusive probability densities for the realization of $n$
jumps at the times $t_1 < t_2 < \ldots < t_{n - 1} < t_n$ and no jumps in
between up to time $t$ {\cite{Barchielli1994a,Budini2013a}}. This
interpretation is justified by making reference to the use of stochastic
master equations in the theory of continuous measurement
{\cite{Barchielli1994a,Barchielli2009}}. Note in particular the crucial fact
that these probability densities do actually depend on the initial state, and
not only on the operators appearing in the master equation. This
interpretation as probability densities is most easily seen considering an
interesting special case. Consider the situation in which the superoperator
$\mathcal{\tilde{J}}$ sends each operator to a fixed statistical operator
$\bar{\rho}$, so that
\begin{eqnarray}
  \mathcal{J} \sigma & = & \bar{\rho} \tmop{Tr} \left( \mathcal{J} \sigma
  \right) .  \label{eq:jfix}
\end{eqnarray}
In this case one has, starting from the state $\bar{\rho}$
\begin{eqnarray}
  \mathpi^n_t \left( t_1, \ldots, t_n \right) & = & \tmop{Tr} \left(
  \mathcal{R} \left( t - t_n \right) \mathcal{J} \ldots \mathcal{J}
  \mathcal{R} \left( t_2 - t_1 \right) \mathcal{J} \mathcal{R} \left( t_1
  \right) \bar{\rho} \right)  \label{eq:pnrenew}\\
  & = & \tmop{Tr} \left( \mathcal{R} \left( t - t_n \right) \mathcal{J}
  \ldots \mathcal{J} \mathcal{R} \left( t_2 - t_1 \right) \bar{\rho} \right)
  \tmop{Tr} \left( \mathcal{J} \mathcal{R} \left( t_1 \right) \bar{\rho}
  \right) \nonumber\\
  & = & \tmop{Tr} \left( \mathcal{R} \left( t - t_n \right) \bar{\rho}
  \right) \ldots \tmop{Tr} \left( \mathcal{J} \mathcal{R} \left( t_2 - t_1
  \right) \bar{\rho} \right) \tmop{Tr} \left( \mathcal{J} \mathcal{R} \left(
  t_1 \right) \bar{\rho} \right) . \nonumber
\end{eqnarray}
Setting
\begin{eqnarray}
  w_0 \left( t \right) & = & \tmop{Tr} \left( \mathcal{} \mathcal{R} \left( t
  \right) \bar{\rho} \right) \nonumber\\
  w \left( t \right) & = & \tmop{Tr} \left( \mathcal{J} \mathcal{R} \left( t
  \right) \bar{\rho} \right), \nonumber
\end{eqnarray}
thanks to the definition of the superoperators Eq.~(\ref{eq:l}) and
Eq.~(\ref{eq:r}) one can immediately check the relation
\begin{eqnarray}
  \frac{\mathd w_0 \left( t \right)}{\tmop{dt}} & = & - \tmop{Tr} \left(
  \mathcal{J} \mathcal{R} \left( t \right) \bar{\rho} \right) \nonumber\\
  & = & - w \left( t \right), \nonumber
\end{eqnarray}
which given the fact that $w_0 \left( t \right)$ has the properties of a
survival probability, the probability of no jumps up to time $t$, implies that
$w \left( t \right)$ is its associated waiting time distribution, the
probability density for a count at time $t$, thus leading to the expression
\begin{eqnarray}
  \mathpi^n_t \left( t_1, \ldots, t_n \right) & = & w_0 \left( t - t_n \right)
  \ldots w \left( t_2 - t_1 \right) w \left( t_1 \right), \nonumber
\end{eqnarray}
which corresponds to a renewal process for the distribution in time of the
jumps {\cite{Ross2007}}.

The expression of the solution of the master equation given by
Eq.~(\ref{eq:physprob}) can be seen as a demixture of the state at time $t$ in
terms of states corresponding to possible trajectories. Each trajectory is
specified by the number and the time of the counts or jumps. The states
associated to the different trajectories are then weighted according to the
probability densities on the trajectory space given by Eq.~(\ref{eq:pn}),
which are determined by the quantum dynamics itself, and therefore provide the
so called physical probabilities. The latter indeed allow to express the
solution of the master equation as average over normalized states arising as
solution of an associated nonlinear stochastic master equation and
corresponding to different trajectories, see
{\cite{Barchielli1994a,Barchielli2009}} for a mathematically more precise
treatment. Alternatively, always exploiting the formalism of stochastic master
equations, one can express the solution of the master equation
Eq.~(\ref{eq:lindblad}) by using as weight an arbitrary reference probability,
independent from the initial state, e.g. a Poisson distribution with a fixed
parameter $\lambda$, so that in this case the probability to have $n$ counts
up to time $t$ at fixed times $t_1 < t_2 < \ldots < t_{n - 1} < t_n$ is
independent of the actual times $\left\{ t_i \right\}_{i = 1, \ldots, n}$ and
is given by $\lambda^n \exp \left( - \lambda t \right)$. In this case however
the state is expressed as demixture with these weights of unnormalized
statistical operators $\tilde{\rho} \left( t \right)$, also called statistical
subcollections, that is positive operators with trace less or equal than one,
which arise as solution of a linear stochastic master equation and take the
form {\cite{Barchielli1994a,Barchielli2009}}
\begin{eqnarray}
  \tilde{\rho} \left( t \right) & = & \lambda^{- n} \exp \left( \lambda t
  \right) \mathcal{R} \left( t - t_n \right) \mathcal{J} \ldots \mathcal{J}
  \mathcal{R} \left( t_2 - t_1 \right) \mathcal{J} \mathcal{R} \left( t_1
  \right) \rho \left( 0 \right),  \label{eq:unnorm}
\end{eqnarray}
leading in the end to the standard expansion of the solution. Knowledge of the
existence of the two alternatives will clarify the nature of the non-Markovian
extension that we shall consider below, which can be seen as arising by
merging the two viewpoints.

Before proceeding let us briefly show how to obtain Eq.~(\ref{eq:physprob})
without the need to resort to the formalism of stochastic master equations for
the statistical operator. Let us start from the expression of the solution in
the familiar form of a Dyson series {\cite{Holevo2001}}
\begin{eqnarray}
  \rho \left( t \right) & = & \mathcal{R} \left( t \right) \rho \left( 0
  \right) + \sum_{n = 1}^{\infty} \int^t_0 \mathd t_n \ldots \int^{t_2}_0
  \mathd t_1  \label{eq:dyson}\\
  &  & \hspace{2em} \times \mathcal{R} \left( t - t_n \right) \mathcal{J}
  \ldots \mathcal{J} \mathcal{R} \left( t_2 - t_1 \right) \mathcal{J}
  \mathcal{R} \left( t_1 \right) \rho \left( 0 \right), \nonumber
\end{eqnarray}
which in particular at variance with Eq.~(\ref{eq:physprob}) immediately shows
linearity and complete positivity of the time evolution. Note further that
according to the given definitions Eq.~(\ref{eq:rtilde}) and Eq.~(\ref{eq:po})
one immediately has the relation
\begin{eqnarray}
  \mathcal{R} \left( t \right) \sigma & = & \tmop{Tr} \left( \mathcal{R}
  \left( t \right) \sigma \right) \frac{\mathcal{R} \left( t \right)
  \sigma}{\tmop{Tr} \left( \mathcal{R} \left( t \right) \sigma \right)}
  \nonumber\\
  & = & \mathpi^0_t \left( t \right) \tilde{\mathcal{R}} \left( t \right)
  \sigma . \nonumber
\end{eqnarray}
Moreover for any superoperator $\mathcal{\tilde{A}}$ homogeneous of order zero
according to Eq.~(\ref{eq:homo}) one can immediately verify the relation
\begin{eqnarray}
  \mathcal{\tilde{A}} \mathcal{R} \left( t - t_n \right) \mathcal{J} \ldots
  \mathcal{J} \mathcal{R} \left( t_1 \right) & = & \mathcal{\tilde{A}}
  \mathcal{} \mathcal{\tilde{\mathcal{R}}} \left( t - t_n \right)
  \mathcal{\tilde{J}} \ldots \mathcal{\tilde{J} }
  \mathcal{\widetilde{\mathcal{R}}} \left( t_1 \right), \nonumber
\end{eqnarray}
so that one has the simple basic relationship
\begin{eqnarray}
  \mathcal{R} \left( t - t_n \right) \mathcal{J} \ldots \mathcal{J}
  \mathcal{R} \left( t_1 \right) \rho \left( 0 \right) & = & \tmop{Tr} \left(
  \mathcal{R} \left( t - t_n \right) \mathcal{J} \ldots \mathcal{J}
  \mathcal{R} \left( t_1 \right) \rho \left( 0 \right) \right) \nonumber\\
  &  & \hspace{1em} \times \frac{\mathcal{R} \left( t - t_n \right)
  \mathcal{J} \ldots \mathcal{J} \mathcal{R} \left( t_1 \right) \rho \left( 0
  \right)}{\tmop{Tr} \left( \mathcal{R} \left( t - t_n \right) \mathcal{J}
  \ldots \mathcal{J} \mathcal{R} \left( t_1 \right) \rho \left( 0 \right)
  \right)} \nonumber\\
  & = & \mathpi^n_t \left( t_1, \ldots, t_n \right)
  \mathcal{\tilde{\mathcal{R}}} \left( t - t_n \right) \mathcal{\tilde{J}}
  \ldots \mathcal{\tilde{J} } \mathcal{\widetilde{\mathcal{R}}} \left( t_1
  \right), \nonumber
\end{eqnarray}
which proves Eq.~(\ref{eq:physprob}). Note that in general the probability
densities $\mathpi^n_t \left( t_1, \ldots, t_n \right)$ do not have any
special properties, apart from being positive and normalized to one when
summed over all $n$ and integrated over all possible intermediate times. As
discussed above a simple situation only appears if the jump operator sends a
generic state to a fixed operator. In this case the probability densities can
be expressed in terms of a unique waiting time distribution.

\section{Collisional models from piecewise dynamics}

In the previous Section we have provided through Eq.~(\ref{eq:physprob}) a
particular representation of the statistical operator solution of a Lindblad
dynamics, which is alternative to the usual Dyson expansion of the solution,
corresponding to Eq.~(\ref{eq:dyson}). Eq.~(\ref{eq:physprob}) is the most
natural starting point to come to a general expression for a collisional
dynamical model. Indeed in a collisional model a dynamics is obtained for a
reduced system by building on three basic quantities. An intercollision time
evolution, which describes the dynamics of the reduced system in between
certain interaction events that can be considered localized in time, a state
transformation described by a quantum channel which describes jumps or events,
the correlation in time between these jumps, which can be described in terms
of the probability density for the jump distribution. Expression
Eq.~(\ref{eq:physprob}) is suggestive in this respect, since all three
elements appear in it. However there is a basic difference in that the
dynamics in between jumps and the effect of the events described as collisions
is not described by linear operators, but rather by the state transformations
Eq.~(\ref{eq:r}) and Eq.~(\ref{eq:l}) respectively. We will however take this
starting point to justify the class of non-Markovian dynamics obtained in
{\cite{Vacchini2013a}}, better elucidating its relationship with the Lindblad
result. This will partially overcome the sudden leap made in
{\cite{Vacchini2013a}} from the Dyson expansion to the generalized master
equation, and explain why the standard Lindblad result is actually only
obtained in a trivial limit.

\begin{figure}[tb]
  \centerline{\includegraphics[width=\columnwidth]{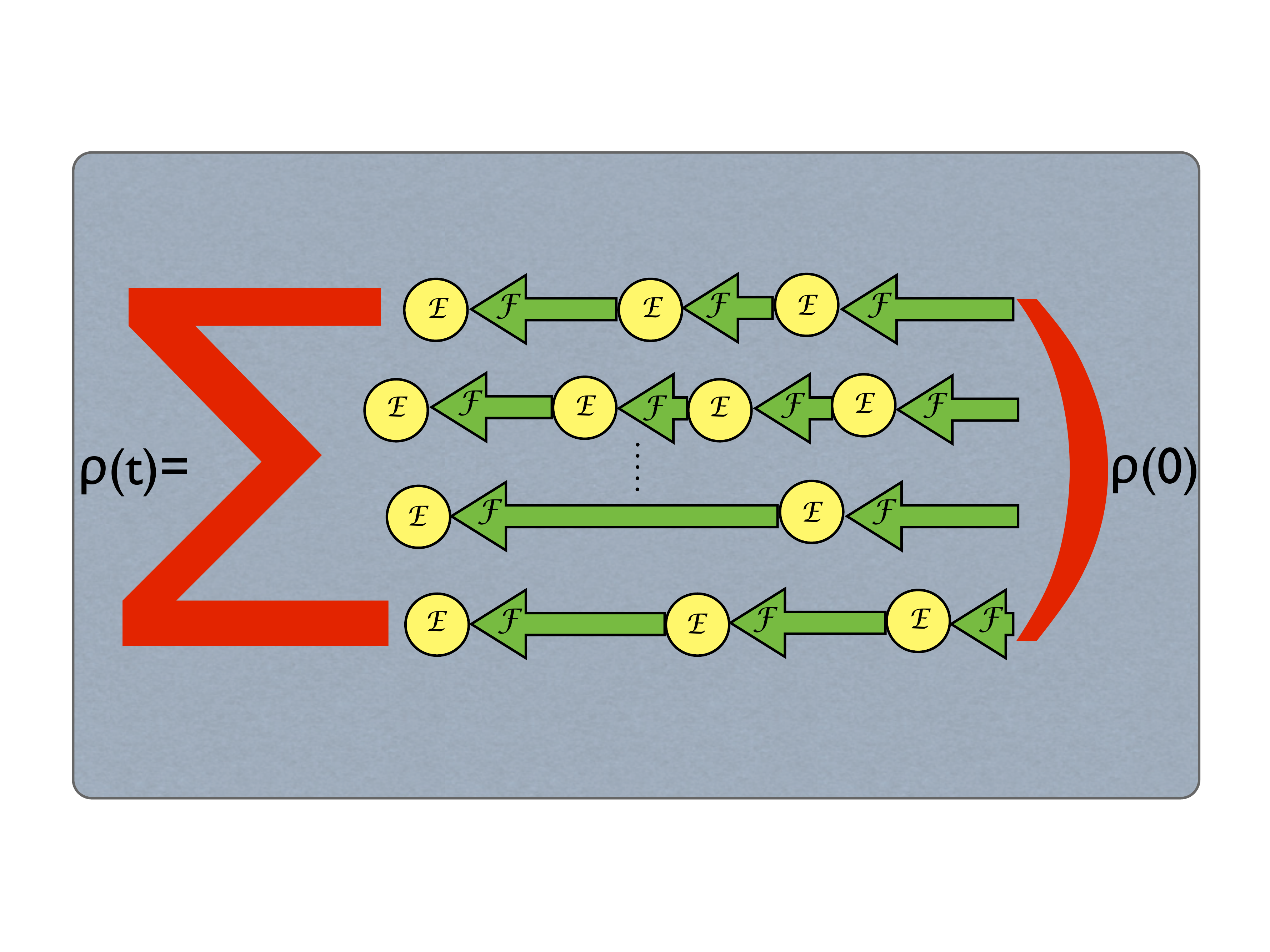}}
  \caption{\label{fig:trajectory} Pictorial representation of the dynamics
  arising from Eq.~(\ref{eq:fint}) according to its trajectory representation
  as given by Eq.~(\ref{eq:nmfinal}). The statistical operator at time $t$
  arises by summing over all trajectories characterized by the repeated action
  of the completely positive trace preserving map $\mathcal{E}$ at jump times
  determined by a fixed waiting time distribution, corresponding to the same
  type of arrows, while the different length of the arrows represents the time
  in between jumps. During this time the system is acted upon by a fixed time
  dependent dynamical map given by $\mathcal{F}$. }
\end{figure}

In Eq.~(\ref{eq:physprob}) the weight of the trajectories, expressed by the so
called physical probability densities, are determined by the operators
describing the dynamics, and the two objects are actually intertwined, as
appears from the fact that in the alternative expression Eq.~(\ref{eq:dyson}),
as well as in the mixture in terms of unnormalized statistical operators given
by Eq.~(\ref{eq:unnorm}), the physical probability densities do not directly
appear. The idea is now to make the probability densities which give the
weight of the trajectories independent from the state transformations between
jumps, as well as from the explicit expression of the jumps operator, thus
introducing an external non trivial distribution of jumps. This is one of the
ingredients in a collisional model. At the same time in order to preserve
linearity and granting complete positivity, we replace the non linear
operators $\mathcal{\tilde{J}}$ and $\mathcal{\tilde{\mathcal{R}}} \left( t
\right)$, with linear completely positive trace preserving transformations,
which provide the other two ingredients of collisional models. Let us
therefore consider the expression
\begin{eqnarray}
  \rho \left( t \right) & = & p_0 \left( t \right) \mathcal{F} \left( t
  \right) \rho \! \left( 0 \right) + \! \sum_{n = 1}^{\infty} \int^t_0 \mathd
  t_n \ldots \int^{t_2}_0 \mathd t_1  \label{eq:nmdyson}\\
  &  & \hspace{2em} \times p_n \left( t_n, \ldots, t_1 \right) \mathcal{F}
  \left( t - t_n \right) \mathcal{E} \ldots \mathcal{E} \mathcal{F} \left( t_2
  - t_1 \right) \mathcal{E} \mathcal{F} \left( t_1 \right) \rho \left( 0
  \right) \nonumber
\end{eqnarray}
where $p_n \left( t_n, \ldots_, t_1 \right)$ denotes the probability density
for the realization of $n$ events up to time $t$, while $\mathcal{E}$ and
$\mathcal{F} \left( t \right)$ denote respectively a completely positive trace
preserving superoperator and a collection of completely positive time
dependent evolutions. Such an expression provides by construction a
realization of a collisional model and realizes a completely positive
transformation on the space of statistical operators. In the general case
however, for a generic weight associated to the different trajectories, that
is a generic distribution of the interaction events, it is not possible to
provide closed evolution equations for the statistical operator of the reduced
system only. This is however the case for a distribution of jumps described by
a renewal process, so that the probability densities read
\begin{eqnarray}
  p_n \left( t_n, \ldots_, t_1 \right) & = & f \left( t_{} - t_n \right)
  \ldots f \left( t_2 - t_1 \right) g \left( t_1 \right),  \label{eq:epd}
\end{eqnarray}
with $f \left( t \right)$ a waiting time distribution, that is a probability
density over the positive reals and $g \left( t \right)$ its associated
survival probability according to the relation
\begin{eqnarray}
  g \left( t \right) & = & 1 - \int^t_0 \mathd \tau f \left( \tau \right) .
  \nonumber
\end{eqnarray}
These relations lead to the expression
\begin{eqnarray}
  \rho \left( t \right) & = & p_0 \left( t \right) \mathcal{F} \left( t
  \right) \rho \! \left( 0 \right) + \! \sum_{n = 1}^{\infty} \int^t_0 \mathd
  t_n \ldots \int^{t_2}_0 \mathd t_1  \label{eq:nmfinal}\\
  &  & \hspace{2em} \times f \left( t_{} - t_n \right) \mathcal{F} \left( t -
  t_n \right) \mathcal{E} \ldots \mathcal{E} f \left( t_2 - t_1 \right)
  \mathcal{F} \left( t_2 - t_1 \right) \mathcal{E} g \left( t_1 \right)
  \mathcal{F} \left( t_1 \right) \rho \left( 0 \right), \nonumber
\end{eqnarray}
schematically depicted in Fig.~\ref{fig:trajectory}
, where a pictorial scheme of the corresponding dynamics is given.
From Eq.~(\ref{eq:nmfinal}) one obtains the integral equation
\begin{eqnarray}
  \rho \left( t \right) & = & g \left( t \right) \mathcal{F} \left( t \right)
  \! + \! \int^t_0 \mathd \tau f \left( t - \tau \right) \mathcal{F} \left( t
  - \tau \right) \mathcal{E} \rho \left( \tau \right),  \label{eq:lt}
\end{eqnarray}
which using the notation $\hat{h} \left( u \right) = \int^{\infty}_0 \mathd t
\mathe^{- ut} h \left( t \right)$ for the Laplace transform, reads
\begin{eqnarray}
  \hat{\rho} \left( u \right) & = & \widehat{g \mathcal{F}} \left( u \right) +
  \widehat{f \mathcal{F}} \left( u \right) \mathcal{E} \hat{\rho} \left( u
  \right) .  \label{eq:lu}
\end{eqnarray}
This expression leads to a formally exact expression for the solution in the
form
\begin{eqnarray}
  \hat{\rho} \left( u \right) & = & \left[ \mathbbm{1} - \widehat{f
  \mathcal{F}} \left( u \right) \mathcal{E} \right]^{- 1} \widehat{g
  \mathcal{F}} \left( u \right) .  \label{eq:sollu}
\end{eqnarray}
Moreover if we start from Eq.~(\ref{eq:lu}), with simple algebra, exploiting
the initial conditions $g \left( 0 \right) = 1$ and $\mathcal{F} \left( 0
\right) = \mathbbm{1}$, one comes to
\begin{eqnarray}
  u \hat{\rho} \left( u \right) - \mathbbm{1} & = & \left[ u \widehat{g
  \mathcal{F}} \left( u \right) - \mathbbm{1} \right] + \left[ u \widehat{f
  \mathcal{F}} \left( u \right) - f \left( 0 \right) \right] \mathcal{E}
  \hat{\rho} \left( u \right) + f \left( 0 \right) \mathcal{E} \hat{\rho}
  \left( u \right), \nonumber
\end{eqnarray}
leading by inversion of the Laplace transform to the closed
integrodifferential equation obeyed by the statistical operator of the reduced
system
\begin{eqnarray}
  \frac{\mathd}{\mathd t} \rho \left( t \right) & = & \int^t_0 \mathd \tau
  \frac{\mathd}{\mathd \left( t - \tau \right)} f \left( t - \tau \right)
  \mathcal{F} \left( t - \tau \right) \mathcal{E} \rho \left( \tau \right) 
  \label{eq:fint}\\
  &  & + f \left( 0 \right) \mathcal{E} \rho \left( t \right) +
  \frac{\mathd}{\mathd t} \left[ g \left( t \right) \mathcal{F} \left( t
  \right) \right] . \nonumber
\end{eqnarray}

As we have shown this result arises building on the representation of the
Lindblad dynamics as given by Eq.~(\ref{eq:physprob}), by substituting the
physical probabilities with a set of probability densities determined by a
single waiting time distribution $f \left( t \right)$ which can be arbitrarily
fixed, and replacing the non linear transformations $\mathcal{\tilde{J}}$ and
$\mathcal{\tilde{\mathcal{R}}} \left( t \right)$, describing a measurement
transformation of the state and strictly connected through Eq.~(\ref{eq:l})
and Eq.~(\ref{eq:r}), with the linear completely positive trace preserving
maps $\mathcal{E}$ and $\mathcal{F} \left( t \right)$ which can be taken
independent of each other. Indeed this changes deeply modify the Lindblad
dynamics, replacing it with a piecewise dynamics characterized by three
independent quantities, so that the distribution of the jumps is not dictated
anymore by the dynamics itself as in Eq.~(\ref{eq:pn}) and conditioned
by the initial state, but rather given by an
external counter. This is reflected by the fact that the Lindblad dynamics is
only recovered in the trivial limit $\mathcal{F} \left( t \right) \rightarrow
\mathe^{t \mathcal{L}}$, with $\mathcal{L}$ a superoperator in Lindblad form
and $\mathcal{E} \rightarrow \mathbbm{1}$, independently of the chosen waiting
time distribution. The most natural interpretation of Eq.~(\ref{eq:fint}) is
therefore as a general scheme of collisional model.

Two important questions related to the obtained completely positive piecewise
dynamics are its degree of non-Markovianity and the possibility to obtain it
as a reduced dynamics from an overall Markovian dynamics in a larger space.
The possible degree of non-Markovianity of these time evolutions has been
discussed in {\cite{Vacchini2013a}}, together with the connection with
different master equations related to collisional models, relying on a
recently introduced notion of non-Markovianity based on the behavior in time
of the distinguishability of different initial reduced states
{\cite{Breuer2009b,Laine2010a}}. The embedding of these dynamics into a
Markovian dynamics in a larger Hilbert space has been most recently addressed
in {\cite{Budini2013a,Budini2013b}}.

\section{Conclusions and outlook}

We have addressed how to formulate a Lindblad dynamics so as to open the way
for the introduction of a general structure of time evolution described by a
collisional model, which allows to consider general non-Markovian dynamics.
This has been obtained by expressing the time evolved statistical operator as
an average over trajectories, weighted by physical probabilities which are
determined by the operators appearing in the Lindblad master equation,
intertwined among them due to probability conservation. Suitably considering
these three elements as independent allows to describe a more general yet
closed piecewise dynamics. This result has opened the way to the study of the
degree of non-Markovianity of the ensuing dynamics, and to the exploration of
their embedding in a Markovian framework in a larger space. It further calls
for microscopic derivations, which could shed light on physically motivated
choices of the waiting time distribution and of the otherwise arbitrary
completely positive trace preserving maps realizing the evolution.

\paragraph*{Acknowledgments.}

The author thanks Prof. Alberto Barchielli for many useful discussions.
Support from COST Action MP 1006 is gratefully acknowledged.

\end{document}